# Classical access structures of ramp secret sharing based on quantum stabilizer codes

Ryutaroh Matsumoto[1,2]

## Abstract

In this paper, we consider to use the quantum stabilizer codes as secret sharing schemes for classical secrets. We give necessary and sufficient conditions for qualified and forbidden sets in terms of quantum stabilizers. Then, we give a Gilbert–Varshamov-type sufficient condition for existence of secret sharing schemes with given parameters, and by using that sufficient condition, we show that roughly 19% of participants can be made forbidden independently of the size of classical secret, in particular when an $n$-bit classical secret is shared among $n$ participants having 1-qubit share each. We also consider how much information is obtained by an intermediate set and express that amount of information in terms of quantum stabilizers. All the results are stated in terms of linear spaces over finite fields associated with the quantum stabilizers.

## 1 Introduction

Secret sharing is a scheme to share a secret among multiple participants so that only *qualified* sets of participants can reconstruct the secret, while *forbidden* sets have no

---

An extended abstract [28] of this manuscript appeared in https://www.lebesgue.fr/content/sem2019-WCC-program the Proceedings of the Eleventh International Workshop on Coding and Cryptography (WCC 2019), Saint-Jacut-de-la-Mer, France, March 31–April 5, 2019. Its presentation slides are available from https://www.slideshare.net/RyutarohMatsumoto. This research is partly supported by JSPS Grant No. 17K06419.

✉ Ryutaroh Matsumoto
 ryutaroh.matsumoto@nagoya-u.jp

1 Department of Information and Communication Engineering, Nagoya University, Nagoya, Japan

2 Department of Mathematical Sciences, Aalborg University, Ålborg, Denmark

information about the secret [36]. A piece of information received by a participant is called a *share*. A set of participants that is neither qualified nor forbidden is said to be *intermediate*. Both secret and shares are traditionally classical information. There exists a close connection between secret sharing and classical error-correcting codes [3,7,10,11,19,23,31].

After the importance of quantum information became well-recognized, secret sharing schemes with quantum shares were proposed [8,15–17,37]. A connection between quantum secret sharing and quantum error-correcting codes has been well known for many years [8,13,15,21,22,35,37], none of which has determined the access structure of secret sharing schemes with classical secrets and quantum shares constructed from quantum stabilizer codes. The well-known classes of quantum error-correcting codes are the CSS codes [6,38], the stabilizer codes [4,5,14] that include the CSS codes as a special case, and their non-binary generalizations [2,18,29].

The access structure of a secret sharing scheme is the set of qualified sets, that of intermediate sets and that of forbidden sets. For practical use of secret sharing, one needs sufficient (and desirably necessary) conditions on qualified sets and forbidden sets. It is natural to investigate access structures of secret sharing schemes constructed from quantum error-correcting codes. For secret sharing schemes with quantum secret and quantum shares, necessary and sufficient conditions for qualified sets and forbidden sets were clarified for the CSS codes [26,37] and the stabilizer codes [13,25]. For classical secret and quantum shares, the access structure was clarified in [26, Section 4.1] with [33, Theorem 1] for the CSS codes but has not been clarified for secret sharing schemes based on quantum stabilizer codes, as far as this author knows.

Advantages of using quantum shares for sharing a classical secret are that we can have smaller size of shares [15, Section 4] and that we can realize access structures that cannot be realized by classical shares [24,27]. For example, it is well known that the size of classical shares cannot be smaller than that of the classical secret in a perfect secret sharing scheme, where *perfect* means that there is no intermediate set, while *ramp* or *non-perfect* means that there exist intermediate sets [39]. On the other hand, the superdense coding can be a secret sharing scheme sharing 2 bits by 2 qubits sent to 2 participants [15, Section 4]. Any participant has no information about the secret, while the 2 participants can reconstruct the secret. We see a perfect threshold scheme sharing 2-bit classical secret by 1-qubit shares. This paper will generalize Gottesman's secret sharing [15, Section 4] to the arbitrary number of participants and the arbitrary size of classical secrets.

In this paper, we give necessary and sufficient conditions for qualified and forbidden sets in terms of the underlying linear spaces over finite fields of quantum stabilizers in Sect. 3, after introducing necessary notations in Sect. 2. Section 3 also includes sufficient conditions in terms of a quantity similar to relative generalized Hamming weight [20] of classical linear codes related to the quantum stabilizers. We also consider how much information is obtained by an intermediate set and express that amount of information in terms of the underlying linear spaces of quantum stabilizers in Sect. 4. Then, we translate our theorems over prime finite fields by the symplectic inner product into arbitrary finite fields, the Euclidean, and the hermitian inner products in Sect. 5. Section 5 also includes an elementary construction by the Reed–Solomon codes as an example of Sect. 5.3. Finally, we give a Gilbert–Varshamov-type sufficient condition for existence of secret sharing schemes with given parameters, and by using that sufficient condition, we show that roughly 19% of participants can be made forbidden independently of the size of classical secret, which cannot be realized by classical shares, in Sect. 6. Concluding remarks are given in Sect. 7. The extended abstract [28] in the workshop had no mathematical proofs and only few examples due to space limitation, and there were confusing typographical errors in the main theorems [28, Theorems 18 and 19].

## 2 Notations

Let $p$ be a prime number, $\mathbf{F}_p$ the finite field with $p$ elements, and $\mathbf{C}_p$ the $p$-dimensional complex linear space. The quantum state space of $n$ qudits is denoted by $\mathbf{C}_p^{\otimes n}$ with its orthonormal basis $\{|\mathbf{v}\rangle : \mathbf{v} \in \mathbf{F}_p^n\}$.

For two vectors $\mathbf{a}, \mathbf{b} \in \mathbf{F}_p^n$, denote by $\langle \mathbf{a}, \mathbf{b} \rangle_E$ the standard Euclidean inner product. For two vectors $(\mathbf{a}|\mathbf{b})$ and $(\mathbf{a}'|\mathbf{b}') \in \mathbf{F}_p^{2n}$, we define the standard symplectic inner product

$$\langle (\mathbf{a}|\mathbf{b}), (\mathbf{a}'|\mathbf{b}') \rangle_s = \langle \mathbf{a}, \mathbf{b}' \rangle_E - \langle \mathbf{a}', \mathbf{b} \rangle_E.$$

For an $\mathbf{F}_p$-linear space $C \subset \mathbf{F}_p^{2n}$, $C^{\perp s}$ denotes its orthogonal space in $\mathbf{F}_p^{2n}$ with respect to $\langle \cdot, \cdot \rangle_s$. Throughout this paper, we always assume $\dim C = n - k$ and $C \subseteq C^{\perp s}$.

For $(\mathbf{a}|\mathbf{b}) \in \mathbf{F}_p^{2n}$, define the $p^n \times p^n$ complex unitary matrix $X(\mathbf{a})Z(\mathbf{b})$ as defined in [18]. An $[[n, k]]_p$ quantum stabilizer codes $Q$ encoding $k$ qudits into $n$ qudits can be defined as a simultaneous eigenspace of all $X(\mathbf{a})Z(\mathbf{b})$ $((\mathbf{a}|\mathbf{b}) \in C)$. Unlike [18], we do not require the eigenvalue of $Q$ to be one.

It is well known in mathematics [1, Chapter 7] that there always exists $C \subseteq C_{\max} \subseteq C^{\perp s}$ such that $C_{\max} = C_{\max}^{\perp s}$. Note that $C_{\max}$ is not unique and usually there are many possible choices of $C_{\max}$. We have $\dim C_{\max} = n$ and have an isomorphism $f : \mathbf{F}_p^k \to C^{\perp s}/C_{\max}$ as linear spaces without inner products. Since $C_{\max} = C_{\max}^{\perp s}$, $C_{\max}$ defines an $[[n, 0]]_p$ quantum stabilizer code $Q_0$. Without loss of generality, we may assume $Q_0 \subset Q$. Let $|\varphi\rangle \in Q_0$ be a quantum state vector. Since $C_{\max} = C_{\max}^{\perp s}$, for a coset $V \in C^{\perp s}/C_{\max}$ and $(\mathbf{a}|\mathbf{b}), (\mathbf{a}'|\mathbf{b}') \in V$, $X(\mathbf{a})Z(\mathbf{b})|\varphi\rangle$ and $X(\mathbf{a}')Z(\mathbf{b}')|\varphi\rangle$ differ by a constant multiple in $\mathbf{C}$ and physically express the same quantum state in $Q$. By an abuse of notation, for a coset $V \in C^{\perp s}/C_{\max}$ we will write $|V\varphi\rangle$ to mean $X(\mathbf{a})Z(\mathbf{b})|\varphi\rangle$ $((\mathbf{a}|\mathbf{b}) \in V)$.

For a given classical secret $\mathbf{m} \in \mathbf{F}_p^k$, we consider the following secret sharing scheme with $n$ participants:

1. $f(\mathbf{m})$ is a coset of $C^{\perp s}/C_{\max}$. Prepare the quantum codeword $|f(\mathbf{m})\varphi\rangle \in Q$ that corresponds to the classical secret $\mathbf{m}$.
2. Distribute each qudit in the quantum codeword $|f(\mathbf{m})\varphi\rangle$ to a participant.

We can also consider a secret sharing scheme for a $k$-qudit secret $|\mathbf{m}\rangle$ with $n$ participants as follows. The reason why we also consider secret sharing schemes with quantum secrets is to contrast the difference between the classical and the quantum access structures of a secret sharing scheme constructed from the same quantum stabilizer in Remarks 7 and 10, while the main focus of the present paper is to share classical secrets.

1. Encode a given quantum secret $\sum_{\mathbf{m}\in\mathbf{F}_p^k} \alpha(\mathbf{m})|\mathbf{m}\rangle$ into the quantum codeword $\sum_{\mathbf{m}\in\mathbf{F}_p^k} \alpha(\mathbf{m})|f(\mathbf{m})\varphi\rangle \in Q$, where $\alpha(\mathbf{m}) \in \mathbf{C}$ are complex coefficients with $\sum_{\mathbf{m}\in\mathbf{F}_p^k} |\alpha(\mathbf{m})|^2 = 1$.
2. Distribute each qudit in the quantum codeword $\sum_{\mathbf{m}\in\mathbf{F}_p^k} \alpha(\mathbf{m})|f(\mathbf{m})\varphi\rangle$ to a participant.

Let $A \subset \{1,\ldots,n\}$ be a set of shares (or equivalently participants), $\overline{A} = \{1,\ldots,n\}\setminus A$, and $\mathrm{Tr}_{\overline{A}}$ the partial trace over $\overline{A}$. For a density matrix $\rho$, $\mathrm{col}(\rho)$ denotes its column space. When $\mathrm{col}(\rho_1), \ldots, \mathrm{col}(\rho_n)$ are orthogonal to each other, that is, $\rho_i \rho_j = 0$ for $i \neq j$, we can distinguish $\rho_1, \ldots, \rho_n$ by a suitable projective measurement with probability 1.

**Definition 1** We say $A$ to be $c$-qualified (classically qualified) if $\mathrm{col}(\mathrm{Tr}_{\overline{A}}(|f(\mathbf{m})\varphi\rangle\langle f(\mathbf{m})\varphi|))$ and $\mathrm{col}(\mathrm{Tr}_{\overline{A}}(|f(\mathbf{m}')\varphi\rangle\langle f(\mathbf{m}')\varphi|))$ are orthogonal to each other for different $\mathbf{m}, \mathbf{m}' \in \mathbf{F}_p^k$. We say $A$ to be $c$-forbidden (classically forbidden) if $\mathrm{Tr}_{\overline{A}}(|f(\mathbf{m})\varphi\rangle\langle f(\mathbf{m})\varphi|)$ is the same density matrix regardless of classical secret $\mathbf{m}$. By a classical access structure, we mean the set of $c$-qualified sets and the set of $c$-forbidden sets.

For a quantum secret, the quantum qualified ($q$-qualified) sets and the quantum forbidden ($q$-forbidden) sets are mathematically defined in [33]. By a quantum access structure, we mean the set of $q$-qualified sets and the set of $q$-forbidden sets.

**Remark 2** When classical shares on $A$ are denoted by $S_A$, the conventional definition of qualifiedness is $I(\mathbf{m}; S_A) = H(\mathbf{m})$ and that of forbiddenness is $I(\mathbf{m}; S_A) = 0$ [39], where $H(\cdot)$ denotes the entropy and $I(\cdot;\cdot)$ denotes the mutual information [9]. Let $\rho_A = \sum_{\mathbf{m}\in\mathbf{F}_p^k} p(\mathbf{m})\mathrm{Tr}_{\overline{A}}(|f(\mathbf{m})\varphi\rangle\langle f(\mathbf{m})\varphi|)$, where $p(\mathbf{m})$ is the probability distribution of classical secrets $\mathbf{m}$. The quantum counterpart of mutual information for classical random variables is the Holevo information $I(\mathbf{m}; \rho_A)$ [32, Section 12.1.1]. $A$ is $c$-qualified if and only if $I(\mathbf{m}; \rho_A) = H(\mathbf{m})$, and is $c$-forbidden if and only if $I(\mathbf{m}; \rho_A) = 0$. Therefore, Definition 1 is a natural generalization of the conventional definition in [39].

**Example 3** We will see how one can express the secret sharing scheme based on superdense coding [15, Section 4] by a quantum stabilizer. Let $p=2$, $n=2$ and $C$ be the zero-dimensional linear space consisting of only the zero vector. Then, $C^{\perp_s} = \mathbf{F}_2^4$. We choose $C_{\max}$ as the space spanned by $(1,1|0,0)$ and $(0,0|1,1)$. For a classical secret $(m_1, m_2) \in \mathbf{F}_2^2$, define the map $f$ as $f(m_1, m_2) = (m_1, 0|m_2, 0) + C_{\max} \in C^{\perp_s}/C_{\max}$. We can choose $[[2,0]]_2$ quantum code $Q_0$ as the one-dimensional complex linear space spanned by the Bell state

$$|\varphi\rangle = \frac{|00\rangle + |11\rangle}{\sqrt{2}},$$

which corresponds to the two-bit secret $(0, 0)$. The secret $(m_1, m_2)$ is encoded to

$$X(m_1, 0)Z(m_2, 0)|\varphi\rangle = \frac{|m_1 0\rangle + (-1)^{m_2}|(1-m_1)1\rangle}{\sqrt{2}}.$$

It is clear that the share set $\{1, 2\}$ is $c$-qualified. When $A = \{1\}$ or $A = \{2\}$, we have

$$\mathrm{Tr}_{\overline{A}}(|f(\mathbf{m})\varphi\rangle\langle f(\mathbf{m})\varphi|) = \frac{1}{2}\begin{pmatrix} 1 & 0 \\ 0 & 1 \end{pmatrix},$$

which means $\{1\}$, $\{2\}$ and $\emptyset$ are $c$-forbidden. We have determined the classical access structure completely, and we see that this scheme is perfect [39] in the sense that there is no intermediate set.

For completeness, we also note its quantum access structure. The set $\{1, 2\}$ is $q$-qualified and $\emptyset$ is $q$-forbidden, of course. By [25, Eq. (3)], we see that $\{1\}$ and $\{2\}$ are intermediate, that is, neither qualified nor forbidden. This quantum access structure exemplifies the fact that $q$-qualifiedness implies $c$-qualifiedness, that $q$-forbiddenness implies $c$-forbiddenness and that their converses are generally false [33, Theorems 1 and 2]. It also exemplifies the fact that if quantum secret is larger than quantum shares, then the scheme cannot be perfect [8,15].

## 3 Necessary and sufficient conditions on classically qualified and classically forbidden sets

Let $A \subset \{1, \ldots, n\}$. Define $\mathbf{F}_p^A = \{(a_1, \ldots, a_n|b_1, \ldots, b_n) \in \mathbf{F}_p^{2n} : (a_i, b_i) = 0$ for $i \notin A\}$. Let $P_A$ be the projection map onto $A$, that is, $P_A(a_1, \ldots, a_n|b_1, \ldots, b_n) = (a_i|b_i)_{i \in A}$.

**Theorem 4** *For the secret sharing scheme described in Sect.* 2, *$A$ is $c$-qualified if and only if*

$$\dim C_{\max}/C = \dim C_{\max} \cap \mathbf{F}_p^A / C \cap \mathbf{F}_p^A. \tag{1}$$

*$A$ is $c$-forbidden if and only if*

$$0 = \dim C_{\max} \cap \mathbf{F}_p^A / C \cap \mathbf{F}_p^A. \tag{2}$$

The proof is given after showing two examples below.

**Example 5** Consider the situation in Example 3. For $A = \{1\}$ or $A = \{2\}$, we see that $C_{\max} \cap \mathbf{F}_2^A$ and $C \cap \mathbf{F}_2^A$ are the zero linear space and that Eq. (2) holds. For $A = \{1, 2\}$, Eq. (1) is clearly true.

**Example 6** In this example, we show that a different choice of $C_{\max}$ gives a different access structure. Let $C$ be as Example 5 and $C_{\max}$ be the linear space generated by $(0, 0|1, 0)$ and $(0, 0|0, 1)$. A classical secret $(m_1, m_2)$ is now encoded to $|m_1 m_2\rangle$. For $A = \{1\}$ or $A = \{2\}$, both (1) and (2) are false and both $A = \{1\}$ and $A = \{2\}$ are intermediate sets. This example shows that the choice of $C_{\max}$ is important.

**Proof** (Theorem 4) Assume Eq. (1). Then, there exists a basis $\{(\mathbf{a}_1|\mathbf{b}_1)+C, \ldots, (\mathbf{a}_k|\mathbf{b}_k) + C\}$ of $C_{\max}/C$ such that $(\mathbf{a}_i|\mathbf{b}_i) \in \mathbf{F}_p^A$. Since $C_{\max}^{\perp s} = C_{\max}$, any two vectors in a coset $V \in C^\perp/C_{\max}$ have the same value of the symplectic inner product against a fixed $(\mathbf{a}_i|\mathbf{b}_i)$, which will be denoted by $\langle(\mathbf{a}_i|\mathbf{b}_i), V\rangle_s$. Suppose that we have two different cosets $V_1, V_2 \in C^\perp/C_{\max}$, and that $\langle(\mathbf{a}_i|\mathbf{b}_i), V_1\rangle_s = \langle(\mathbf{a}_i|\mathbf{b}_i), V_2\rangle_s$ for all $i$. Since $C_{\max}^{\perp s} = C_{\max}$, it means that $V_1 - V_2 = C_{\max}$ is zero in $C^\perp/C_{\max}$, a contradiction. We have seen that any two different cosets have different symplectic inner product values against some $(\mathbf{a}_i|\mathbf{b}_i)$. For each $i$, the $n$ participants can collectively perform quantum projective measurement corresponding to the eigenspaces of $X(\mathbf{a}_i)Z(\mathbf{b}_i)$ and can determine the symplectic inner product[1] $\langle(\mathbf{a}_i|\mathbf{b}_i), f(\mathbf{m})\rangle_s$ as [18, Lemma 5] when the classical secret is $\mathbf{m}$. Since $(\mathbf{a}_i|\mathbf{b}_i)$ has nonzero components only at $A$, the above measurement can be done only by $A$, which means $A$ can reconstruct $\mathbf{m}$.

Assume that Eq. (1) is false. Since the orthogonal space of $C$ in $\mathbf{F}_p^A$ is isomorphic to $P_A(C^{\perp s})$, which can be seen as the almost same argument as the duality between shortened linear codes and punctured linear codes [34], we see that $\dim P_A(C^{\perp s})/P_A(C_{\max}) < \dim C^{\perp s}/C_{\max}$. This means that there exists two different classical secrets $\mathbf{m}_1$ and $\mathbf{m}_2$ such that $P_A(f(\mathbf{m}_1)) = P_A(f(\mathbf{m}_2))$. This means that the encoding procedures of $\mathbf{m}_1$ and $\mathbf{m}_2$ are exactly same on $A$ and produce the same density matrix on $A$, which shows that $A$ is not $c$-qualified.

Assume Eq. (2). Then, we have $\dim P_A(C^{\perp s})/P_A(C_{\max}) = 0$. This means that for all classical secrets $\mathbf{m}$, $P_A(f(\mathbf{m}))$ and their encoding procedures on $A$ are same, which produces the same density matrix on $A$ regardless of $\mathbf{m}$. This shows that $A$ is $c$-forbidden.

Assume that Eq. (2) is false. Then, there exist two different classical secrets $\mathbf{m}_1$, $\mathbf{m}_2$, and $(\mathbf{a}|\mathbf{b}) \in C_{\max} \cap \mathbf{F}_p^A \setminus C \cap \mathbf{F}_p^A$ such that

$$\langle(\mathbf{a}|\mathbf{b}), f(\mathbf{m}_1)\rangle_s \neq \langle(\mathbf{a}|\mathbf{b}), f(\mathbf{m}_2)\rangle_s.$$

By [18, Lemma 5], this means that the quantum measurement corresponding to $X(\mathbf{a})Z(\mathbf{b})$ gives different outcomes with $\mathrm{Tr}_{\overline{A}}(|f(\mathbf{m}_1)\varphi\rangle\langle f(\mathbf{m}_1)\varphi|)$ and $\mathrm{Tr}_{\overline{A}}(|f(\mathbf{m}_2)\varphi\rangle\langle f(\mathbf{m}_2)\varphi|)$. Since $(\mathbf{a}|\mathbf{b}) \in \mathbf{F}_p^A$, measurement of $X(\mathbf{a})Z(\mathbf{b})$ can be performed only by participants in $A$. These observations show that $A$ is not $c$-forbidden. □

**Remark 7** A necessary and sufficient condition for $A$ being $q$-qualified is [25, Eq. (3)]

$$C^{\perp s} \cap \mathbf{F}_p^{\overline{A}} = C_{\max} \cap \mathbf{F}_p^{\overline{A}} = C \cap \mathbf{F}_p^{\overline{A}}. \tag{3}$$

Since $\ker(P_A) = \mathbf{F}_p^{\overline{A}}$, we have $\dim P_A(C^{\perp s})/P_A(C_{\max}) = k$. The relation between duals of punctured codes and shortened codes [34] implies $\dim C_{\max} \cap \mathbf{F}_p^A/C \cap \mathbf{F}_p^A = k$. Therefore, Eq. (3) implies Eq. (1).

---

[1] If we assume a non-prime finite field $\mathbf{F}_q$ as our base field, then the quantum measurement outcome just determines [18, Lemma 5] $\mathrm{Tr}_{q/p}(\langle(\mathbf{a}_i|\mathbf{b}_i), f(\mathbf{m})\rangle_s)$ in place of $\langle(\mathbf{a}_i|\mathbf{b}_i), f(\mathbf{m})\rangle_s$, where $\mathrm{Tr}_{q/p}$ is the trace map from $\mathbf{F}_q$ to its prime subfield $\mathbf{F}_p$. Assuming a non-prime field $\mathbf{F}_q$ significantly complicates the proofs of Theorem 4 and Lemma 11. So we assume a prime finite field until Sect. 5.

Similarly, by [15, Corollary 2], necessary and sufficient condition for $A$ being $q$-forbidden is

$$C^{\perp_s} \cap \mathbf{F}_p^A = C_{\max} \cap \mathbf{F}_p^A = C \cap \mathbf{F}_p^A. \qquad (4)$$

By a similar argument, we see that Eq. (4) implies Eq. (2).

Next, we give sufficient conditions in terms of the coset distance [11] or the first relative generalized Hamming weight [20]. To do so, we have to slightly modify them. For $(\mathbf{a}|\mathbf{b}) = (a_1, \ldots, a_n|b_1, \ldots, b_n) \in \mathbf{F}_p^n$, define its symplectic weight $\mathrm{swt}(\mathbf{a}|\mathbf{b}) = |\{i : (a_i, b_i) \neq (0, 0)\}|$. For $V_2 \subset V_1 \subset \mathbf{F}_p^{2n}$, we define their coset distance as $d_s(V_1, V_2) = \min\{\mathrm{swt}(\mathbf{a}|\mathbf{b}) : (\mathbf{a}|\mathbf{b}) \in V_1 \setminus V_2\}$.

**Theorem 8** *If $|A| \leq d_s(C_{\max}, C) - 1$, then $A$ is c-forbidden. If $|A| \geq n - d_s(C^{\perp_s}, C_{\max}) + 1$, then $A$ is c-qualified.*

**Example 9** Consider the situation in Example 5. We have $d_s(C^\perp, C_{\max}) = 1$, which implies that 2 shares form a $c$-qualified set. We also have $d_s(C_{\max}, C) = 2$, which implies that 1 share forms a $c$-forbidden set.

**Proof** (Theorem 8) If $|A| \leq d_s(C_{\max}, C) - 1$, then there is no $(\mathbf{a}|\mathbf{b}) \in C_{\max} \cap \mathbf{F}_p^A \setminus C \cap \mathbf{F}_p^A$ and Eq. (2) holds.

Assume that $|A| \geq n - d_s(C^{\perp_s}, C_{\max}) + 1$, or equivalently, $|\overline{A}| \leq d_s(C^{\perp_s}, C_{\max}) - 1$. We have $C^{\perp_s} \cap \mathbf{F}_p^{\overline{A}} = C_{\max} \cap \mathbf{F}_p^{\overline{A}}$. We also have $\mathbf{F}_p^{\overline{A}} = \ker(P_A)$, which means $\dim P_A(C^{\perp_s}) - \dim P_A(C_{\max}) = \dim C^{\perp_s} - \dim C_{\max} = k$. Since $\dim C_{\max} \cap \mathbf{F}_p^A - \dim C \cap \mathbf{F}_p^A = \dim P_A(C^{\perp_s}) - \dim P_A(C_{\max}) = k$, we see that Eq. (1) holds with $A$. □

**Remark 10** By Remark 7 and a similar argument to the last proof, we see that if $|A| \leq d_s(C^{\perp_s}, C) - 1$, then $A$ is $q$-forbidden and that if $|A| \geq n - d_s(C^{\perp_s}, C) + 1$, then $A$ is $q$-qualified. Note that these observations can also be deduced from quantum erasure decoding and [15, Corollary 2] and are not novel.

## 4 Amount of information possessed by an intermediate set

Let $A \subset \{1, \ldots, n\}$ with $A \neq \emptyset$ and $A \neq \{1, \ldots, n\}$. In this section, we study the amount of information possessed by $A$.

**Lemma 11** *For two classical secrets $\mathbf{m}_1$ and $\mathbf{m}_2$, we have*

- $\mathrm{Tr}_{\overline{A}}(|f(\mathbf{m}_1)\varphi\rangle\langle f(\mathbf{m}_1)\varphi|) = \mathrm{Tr}_{\overline{A}}(|f(\mathbf{m}_2)\varphi\rangle\langle f(\mathbf{m}_2)\varphi|)$ *if and only if $f(\mathbf{m}_1)$ and $f(\mathbf{m}_2)$ give the same symplectic inner product for all vectors in $C_{\max} \cap \mathbf{F}_p^A$, and*
- $\mathrm{col}(\mathrm{Tr}_{\overline{A}}(|f(\mathbf{m}_1)\varphi\rangle\langle f(\mathbf{m}_1)\varphi|))$ *and* $\mathrm{col}(\mathrm{Tr}_{\overline{A}}(|f(\mathbf{m}_2)\varphi\rangle\langle f(\mathbf{m}_2)\varphi|))$ *are orthogonal to each other if and only if $f(\mathbf{m}_1)$ and $f(\mathbf{m}_2)$ give different symplectic inner products for some vector $(\mathbf{a}|\mathbf{b})$ in $C_{\max} \cap \mathbf{F}_p^A$.*

**Proof** Assume that $f(\mathbf{m}_1)$ and $f(\mathbf{m}_2)$ give the same symplectic inner product for all vectors in $C_{\max} \cap \mathbf{F}_p^A$. Then, we have $\{P_A(\mathbf{a}|\mathbf{b}) : (\mathbf{a}|\mathbf{b}) \in f(\mathbf{m}_1)\} = \{P_A(\mathbf{a}|\mathbf{b}) :$

$(\mathbf{a}|\mathbf{b}) \in f(\mathbf{m}_2)\}$, and the encoding procedure on $A$ is the same for $\mathbf{m}_1$ and $\mathbf{m}_2$, which shows $\text{Tr}_{\overline{A}}(|f(\mathbf{m}_1)\varphi\rangle\langle f(\mathbf{m}_1)\varphi|) = \text{Tr}_{\overline{A}}(|f(\mathbf{m}_2)\varphi\rangle\langle f(\mathbf{m}_2)\varphi|)$.

Assume that $f(\mathbf{m}_1)$ and $f(\mathbf{m}_2)$ give different symplectic inner products for some vector $(\mathbf{a}|\mathbf{b})$ in $C_{\max} \cap \mathbf{F}_p^A$. Then, the quantum measurement corresponding to $X(\mathbf{a})Z(\mathbf{b})$ can be performed only by the participants in $A$ and by [18, Lemma 5] the outcomes for $|f(\mathbf{m}_1)\varphi\rangle$ and $|f(\mathbf{m}_2)\varphi\rangle$ are different with probability 1. This means that $\text{col}(\text{Tr}_{\overline{A}}(|f(\mathbf{m}_1)\varphi\rangle\langle f(\mathbf{m}_1)\varphi|))$ and $\text{col}(\text{Tr}_{\overline{A}}(|f(\mathbf{m}_2)\varphi\rangle\langle f(\mathbf{m}_2)\varphi|))$ are orthogonal to each other. □

**Proposition 12** *If* $\dim C_{\max} \cap \mathbf{F}_p^A / C \cap \mathbf{F}_p^A = \ell$, *then the number of density matrices in* $\Lambda = \{\text{Tr}_{\overline{A}}(|f(\mathbf{m})\varphi\rangle\langle f(\mathbf{m})\varphi|) : \mathbf{m} \in \mathbf{F}_p^k\}$ *is* $p^\ell$.

*For a fixed density matrix* $\rho \in \Lambda$, *the number of classical secrets* $\mathbf{m}$ *such that* $\rho = \text{Tr}_{\overline{A}}(|f(\mathbf{m})\varphi\rangle\langle f(\mathbf{m})\varphi|)$ *is exactly* $p^{k-\ell}$.

*Proof* If $P_A(\mathbf{u}_1|\mathbf{v}_1) + P_A(C_{\max}) \neq P_A(\mathbf{u}_2|\mathbf{v}_2) + P_A(C_{\max})$ for $(\mathbf{u}_i|\mathbf{v}_i) \in f(\mathbf{m}_i)$ with classical secrets $\mathbf{m}_i$ ($i = 1, 2$), then by Lemma 11 $\text{col}(\text{Tr}_{\overline{A}}(|f(\mathbf{m}_1)\varphi\rangle\langle f(\mathbf{m}_1)\varphi|))$ and $\text{col}(\text{Tr}_{\overline{A}}(|f(\mathbf{m}_2)\varphi\rangle\langle f(\mathbf{m}_2)\varphi|))$ are orthogonal. By the assumption, we have $\dim C_{\max} \cap \mathbf{F}_p^A / C \cap \mathbf{F}_p^A = \dim P_A(C^{\perp s})/P_A(C_{\max}) = \ell$. There are $p^\ell$ elements in $P_A(C^{\perp s})/P_A(C_{\max})$, which shows the first claim.

The composite $\mathbf{F}_p$-linear map "mod $P_A(C_{\max})$" $\circ P_A \circ f$ from $\mathbf{F}_p^k$ to $P_A(C^{\perp s})/P_A(C_{\max})$ is surjective. Thus, the dimension of its kernel is $k - \ell$, which shows the second claim. □

**Definition 13** In light of Proposition 12, the amount of information possessed by a set $A$ of participants is defined as

$$(\log_2 p) \times \dim C_{\max} \cap \mathbf{F}_p^A / C \cap \mathbf{F}_p^A. \quad (5)$$

*Remark 14* When the probability distribution of classical secrets $\mathbf{m}$ is uniform, the quantity in Definition 13 is equal to the Holevo information [32, Section 12.1.1] counted in $\log_2$. To see this, firstly, the set $\Lambda$ in Proposition 12 consists of non-overlapping projection matrices and each matrix commutes with every other matrices in $\Lambda$. So the Holevo information is just equal to the classical mutual information [9] between random variable $X$, corresponding to classical secrets in $\mathbf{F}_p^k$, and random variable $Y$, corresponding to matrices in $\Lambda$, where $Y$ is given as a surjective function of $X$. By Proposition 12, $Y$ has the uniform probability distribution. Therefore, $I(X; Y) = H(Y) = \log_2 |\Lambda| = $ Eq. (5).

We say that a secret sharing scheme is $r_i$-reconstructible if $|A| \geq r_i$ implies $A$ has $i \log_2 p$ or more bits of information [12]. We say that a secret sharing scheme is $t_i$-private if $|A| \leq t_i$ implies $A$ has less than $i \log_2 p$ bits of information [12]. In order to express $r_i$ and $t_i$ in terms of combinatorial properties of $C$, we introduce a slightly modified version of the relative generalized Hamming weight [20].

**Definition 15** For two linear spaces $V_2 \subset V_1 \subset \mathbf{F}_p^{2n}$ and $i = 1, \ldots, k$, define the $i$th relative generalized symplectic weight

$$d_s^i(V_1, V_2) = \min\{|A| : \dim \mathbf{F}_p^A \cap V_1 - \dim \mathbf{F}_p^A \cap V_2 \geq i\}. \quad (6)$$

Note that $d_s^1 = d_s$. The following theorem generalizes Theorem 8.

**Theorem 16**

$$t_i \geq d_s^i(C_{\max}, C) - 1,$$
$$r_{k+1-i} \leq n - d_s^i(C^{\perp s}, C_{\max}) + 1.$$

**Example 17** Consider the situation of Example 9. We have $d_s^1(C_{\max}, C) = d_s^2(C_{\max}, C) = 2$, and $d_s^1(C^{\perp s}, C_{\max}) = d_s^2(C^{\perp s}, C_{\max}) = 1$. Unlike the relative generalized Hamming weight, we do not have the strict monotonicity in $i$ of $d_s^i$.

*Proof* (Theorem 16) Assume that $|A| \leq t_i$. By definition of $d_s^i$, $\dim C_{\max} \cap \mathbf{F}_p^A / C \cap \mathbf{F}_p^A \leq i - 1$, which shows the first claim.

Assume that $|A| \geq r_i$. Then, $|\overline{A}| \leq d_s^i(C^{\perp s}, C_{\max}) - 1$, which implies $\dim C^{\perp s} \cap \mathbf{F}_p^{\overline{A}} / C_{\max} \cap \mathbf{F}_p^{\overline{A}} \leq i - 1$. The last inequality implies $\dim C_{\max} \cap \mathbf{F}_p^A / C \cap \mathbf{F}_p^A \geq k - i + 1$. which shows the second claim. □

## 5 Translations to arbitrary finite fields and to the ordinary Hamming weight

### 5.1 Translation to arbitrary finite fields

Let $q = p^\mu$ with $\mu \geq 1$, and $\{\gamma_1, \ldots, \gamma_\mu\}$ be a fixed $\mathbf{F}_p$-basis of $\mathbf{F}_q$. Ashikhmin and Knill [2] proposed the following translation from $\mathbf{F}_q$ to $\mathbf{F}_p$ for quantum stabilizer codes. Let $M$ be a $\mu \times \mu$ invertible matrix over $\mathbf{F}_p$ whose $(i, j)$ element is $\mathrm{Tr}_{q/p}(\gamma_i \gamma_j)$, where $\mathrm{Tr}_{q/p}$ is the trace map from $\mathbf{F}_q$ to $\mathbf{F}_p$. Let $\phi$ be an $\mathbf{F}_p$-linear isomorphism sending $(a_{1,1}, \ldots, a_{1,\mu}, a_{2,1}, \ldots, a_{n,\mu} | b_{1,1}, \ldots, b_{1,\mu}, b_{2,1}, \ldots, b_{n,\mu}) \in \mathbf{F}_p^{2\mu n}$ to

$$\left( \sum_{j=1}^\mu a_{1,j}\gamma_j, \ldots, \sum_{j=1}^\mu a_{n,j}\gamma_j \,\middle|\, \sum_{j=1}^\mu b'_{1,j}\gamma_j, \ldots, \sum_{j=1}^\mu b'_{n,j}\gamma_j \right) \in \mathbf{F}_q^{2n},$$

where $(b'_{i,1}, \ldots, b'_{i,\mu}) = (b_{i,1}, \ldots, b_{i,\mu})M^{-1}$ for $i = 1, \ldots, n$.

Ashikhmin and Knill proved the following.

**Proposition 18** [2] *Let* $C \subset \mathbf{F}_q^{2n}$. *Then,* $\dim_{\mathbf{F}_p} \phi^{-1}(C) = \mu \dim_{\mathbf{F}_q} C$, *and* $\phi^{-1}(C)^{\perp s} = \phi^{-1}(C^{\perp s})$, *where* $\dim_{\mathbf{F}_q}$ *is the dimension of a linear space considered over* $\mathbf{F}_q$.

Let $C \subset C_{\max} = C_{\max}^{\perp s} \subset C^{\perp s} \subset \mathbf{F}_q^{2n}$ with $\dim_{\mathbf{F}_q} C = n - k$. Then, we have $\phi^{-1}(C) \subset \phi^{-1}(C_{\max}) = \phi^{-1}(C_{\max})^{\perp s} \subset \phi^{-1}(C)^{\perp s} \subset \mathbf{F}_p^{2\mu n}$ and we can construct a secret sharing scheme by $\phi^{-1}(C) \subset \phi^{-1}(C_{\max})$. It encodes $k\mu \log_2 p = k \log_2 q$ bits of classical secrets $\mathbf{m} \in \mathbf{F}_q^k$ into $\mu n$ qudits in $\mathbf{C}_p$, which can also be seen as $n$ qudits in $\mathbf{C}_q$, where $\mathbf{C}_q$ is the $q$-dimensional complex linear space. Let $A \subset \{1, \ldots, n\}$.

By abuse of notation, by $\mathbf{F}_p^A$ we mean $\{(a_{1,1}, \ldots, a_{1,\mu}, a_{2,1}, \ldots, a_{n,\mu} | b_{1,1}, \ldots, b_{1,\mu}, b_{2,1}, \ldots, b_{n,\mu}) \in \mathbf{F}_p^{2\mu n} : a_{i,j} = b_{i,j} = 0 \text{ for } i \notin A \text{ and } j = 1, \ldots, \mu\}$.

We consider each qudit in $\mathbf{C}_q$ of the quantum codeword as a share and examine the property of a share set $A$. We have

$$\dim_{\mathbf{F}_q} C_{\max} \cap \mathbf{F}_q^A / C \cap \mathbf{F}_q^A = \mu \dim_{\mathbf{F}_p} \phi^{-1}(C_{\max}) \cap \mathbf{F}_p^A / \phi^{-1}(C) \cap \mathbf{F}_p^A. \quad (7)$$

Equation (7) together with Theorem 4 implies

- $A$ is qualified if and only if $\dim_{\mathbf{F}_q} C_{\max} \cap \mathbf{F}_q^A / C \cap \mathbf{F}_q^A = \dim_{\mathbf{F}_q} C_{\max}/C$, and
- $A$ is forbidden if and only if $\dim_{\mathbf{F}_q} C_{\max} \cap \mathbf{F}_q^A / C \cap \mathbf{F}_q^A = 0$.

The above observation shows that Theorems 4 and 8 also hold for $\mathbf{F}_q$. In addition, Eq. (7) means that a share set $A$ has ($\log_2 q \times \dim_{\mathbf{F}_q} C_{\max} \cap \mathbf{F}_q^A / C \cap \mathbf{F}_q^A$)-bits of information about the secret $\mathbf{m} \in \mathbf{F}_q^k$, also generalizes the proof argument of Theorem 16, and implies that Theorem 16 also holds for $\mathbf{F}_q$. In the sequel, we consider a qudit in $\mathbf{C}_q$ as each share and dim means the dimension over $\mathbf{F}_q$.

### 5.2 Translation to the Hamming distance and the hermitian inner product

Many of results in the symplectic construction of quantum error-correcting codes over $\mathbf{F}_q$ are translated to $\mathbf{F}_{q^2}$-linear codes with the hermitian inner product [2,18,29]. For $\mathbf{x} \in \mathbf{F}_{q^2}^n$ define $\mathbf{x}^q$ as the component-wise $q$th power of $\mathbf{x}$. For two vectors $\mathbf{x}, \mathbf{y} \in \mathbf{F}_{q^2}$, define the hermitian inner product as $\langle \mathbf{x}, \mathbf{y} \rangle_h = \langle \mathbf{x}^q, \mathbf{y} \rangle_E$. For $D \subset \mathbf{F}_{q^2}^n$, $D^{\perp h}$ denotes the orthogonal space of $D$ with respect to the hermitian inner product.

Only in Sects. 5.2, 5.3 and 5.4, for $A \subset \{1, \ldots, n\}$, define $\mathbf{F}_q^A = \{(a_1, \ldots, a_n) \in \mathbf{F}_q^n : a_i = 0 \text{ for } i \notin A\}$, and define $P_A$ to be the projection map onto $A$, that is, $P_A(a_1, \ldots, a_n) = (a_i)_{i \in A}$.

**Theorem 19** *Let $D \subset \mathbf{F}_{q^2}^n$ be an $\mathbf{F}_{q^2}$-linear space. We assume $\dim D = k'$ and there exists $D_{\max}$ such that $D \subset D_{\max} \subset D^{\perp h}$ and $D_{\max} = D_{\max}^{\perp h}$, which implies $\dim D_{\max} = n/2$. Then, $D$ defines a secret sharing scheme based on the quantum stabilizer defined by $D$ encoding $n - 2k'$ symbols in $\mathbf{F}_q$. A set $A \subset \{1, \ldots, n\}$ is c-qualified if and only if $\dim D_{\max}/D = \dim D_{\max} \cap \mathbf{F}_{q^2}^A / D \cap \mathbf{F}_{q^2}^A$. A set $A \subset \{1, \ldots, n\}$ is c-forbidden if and only if $0 = \dim D_{\max} \cap \mathbf{F}_{q^2}^A / D \cap \mathbf{F}_{q^2}^A$. If $|A| \geq n - d_H(D^{\perp h}, D_{\max}) + 1$, then $A$ is c-qualified, and if $|A| \leq d_H(D_{\max}, D) - 1$, then $A$ is c-forbidden, where $d_H$ is the coset distance [11], or equivalently, the first relative generalized Hamming weight [20].*

*Proof* The proof is almost same as [18]. □

**Example 20** Consider the situation in Example 9. Then, $D = \{0\}$ and $D_{\max}$ is the one-dimensional $\mathbf{F}_4$-linear space spanned by $(1, 1)$.

### 5.3 Translation to the Hamming distance and the Euclidean inner product

Let $C_2 \subset C_1 \subset \mathbf{F}_q^n$. A method to construct symplectic-self-orthogonal $C \subset \mathbf{F}_q^{2n}$ is to use $\{(\mathbf{a}|\mathbf{b}) : \mathbf{a} \in C_2, \mathbf{b} \in C_1^{\perp_E}\}$ as $C$ [5,18], where "$\perp E$" denotes the Euclidean dual. We have $C^{\perp_s} = \{(\mathbf{a}|\mathbf{b}) : \mathbf{a} \in C_1, \mathbf{b} \in C_2^{\perp_E}\}$.

**Example 21** Example 6 can also be described by $C_2 = \{0\}$, $C_1 = \mathbf{F}_2^2$, and $C'_{\max} = \{(\mathbf{a}|\mathbf{b}) : \mathbf{a} \in C_2, \mathbf{b} \in C_2^{\perp_E}\}$.

**Remark 22** A suitable choice of $C_{\max}$ is unclear as of this writing. A valid choice is $C'_{\max} = \{(\mathbf{a}|\mathbf{b}) : \mathbf{a} \in C_2, \mathbf{b} \in C_2^{\perp_E}\}$, which gives the standard encoding [6,38] of the CSS codes. But this choice gives no advantage over the purely classical secret sharing constructed from linear codes $C_2 \subset C_1$ [3,7,19,23]. Because the necessary and sufficient condition for c-qualified $A$ is $\dim P_A(C_1)/P_A(C_2) = \dim C_1/C_2$ and the necessary and sufficient condition for c-forbidden $A$ is $\dim P_A(C_1)/P_A(C_2) = 0$ by combining [26, Section 4.1] and [33, Theorem 1], which are exactly same [12] as those of the purely classical secret sharing constructed from $C_2 \subset C_1$.

**Theorem 23** *Let $E \subset \mathbf{F}_q^n$ be the $\mathbf{F}_q$-linear space. We assume $\dim E = k'$, and there exists $E_{\max}$ such that $E \subset E_{\max} \subset E^{\perp_E}$ and $E_{\max} = E_{\max}^{\perp_E}$, which implies $\dim E_{\max} = n/2$. Then, $E$ defines a secret sharing scheme based on the quantum stabilizer defined by $E$ encoding $n - 2k'$ symbols in $\mathbf{F}_q$. A set $A \subset \{1, \ldots, n\}$ is c-qualified if and only if $\dim E_{\max}/E = \dim E_{\max} \cap \mathbf{F}_q^A / E \cap \mathbf{F}_q^A$. A set $A \subset \{1, \ldots, n\}$ is c-forbidden if and only if $0 = \dim E_{\max} \cap \mathbf{F}_q^A / E \cap \mathbf{F}_q^A$. If $|A| \geq n - d_H(E^{\perp_E}, E_{\max}) + 1$, then $A$ is c-qualified, and if $|A| \leq d_H(E_{\max}, E) - 1$, then $A$ is c-forbidden.*

*Proof* The proof is almost same as [18]. □

**Example 24** Example 3 is restored by choosing $E = \{0\}$, $E^{\perp_E} = \mathbf{F}_2^2$, and $E_{\max}$ as the $\mathbf{F}_2$-linear space spanned by $(1, 1)$. Thus, we see that Theorem 23, in contrast to Remark 22, can provide a secret sharing scheme with an advantage over purely classical secret sharing.

### 5.4 Construction by the Reed–Solomon codes

Also as an example of Theorem 23, an elementary construction by the Reed–Solomon (RS) codes will be shown below. In Sect. 5.4, assume that $n = q$ and $k$ are positive even integers. Fix $n$ distinct elements $\alpha_1, \ldots, \alpha_n \in \mathbf{F}_q$. Define

$$\mathrm{RS}(n, k) = \{(g(\alpha_1), \ldots, g(\alpha_n)) : g(x) \in \mathbf{F}_q[x], \deg g(x) < k\}.$$

Then, $\mathrm{RS}(n, k)^{\perp_E} = \mathrm{RS}(n, n - k)$ as $n = q$.

For a linear space $V \subset \mathbf{F}_q^n$, its $j$th generalized Hamming weight $d_H^j(V)$ is defined by [34]

$$d_H^j(V) = \min\{|A| : \dim \mathbf{F}_q^A \cap V \geq j\}.$$

For RS codes, $d_H^j(\text{RS}(n, k)) = n - k + j$ [34].

Define
$$C = \{(\mathbf{a}|\mathbf{b}) : \mathbf{a}, \mathbf{b} \in \text{RS}(n, (n-k)/2)\} \subset \mathbf{F}_q^{2n},$$
$$C_{\max} = \{(\mathbf{a}|\mathbf{b}) : \mathbf{a}, \mathbf{b} \in \text{RS}(n, n/2)\} \subset \mathbf{F}_q^{2n},$$

which correspond to $E = \text{RS}(n, (n-k)/2)$ and $E_{\max} = \text{RS}(n, n/2)$ in Theorem 23. Then, we have

$$C_{\max} = C_{\max}^{\perp s},$$
$$C^{\perp s} = \{(\mathbf{a}|\mathbf{b}) : \mathbf{a}, \mathbf{b} \in \text{RS}(n, (n+k)/2)\},$$
$$\dim C = n - k,$$
$$d_s^j(C_{\max}, C) \geq d_s^j(C_{\max}, \{\mathbf{0}\}) \geq d_H^{\lceil j/2 \rceil}(\text{RS}(n, n/2)) = \left\lceil \frac{n+j}{2} \right\rceil, \quad (8)$$
$$d_s^j(C^{\perp s}, C_{\max}) \geq d_s^j(C^{\perp s}, \{\mathbf{0}\}) \geq d_H^{\lceil j/2 \rceil}(\text{RS}(n, (n+k)/2)) = \left\lceil \frac{n-k+j}{2} \right\rceil. \quad (9)$$

In particular, we have $d_s(C_{\max}, C) \geq n/2 + 1$ and $d_s(C^{\perp s}, C_{\max}) \geq (n-k)/2 + 1$. For a set of participants $A \subseteq \{1, \ldots, n\}$, $A$ is forbidden if $|A| \leq n/2$ and $A$ is qualified if $|A| \geq (n+k)/2$, by Theorem 8 or also by Theorem 23.

By Theorem 16, $t_i \geq \lceil (n+i)/2 \rceil - 1$ and $r_i \leq \lceil (n+i)/2 \rceil$. In addition, by the definitions of $r_i$ and $t_i$, we have $r_i \geq t_i + 1$. So we see that $t_i = \lceil (n+i)/2 \rceil - 1$ and $r_i = \lceil (n+i)/2 \rceil$, which also imply that inequalities (8) and (9) are in fact equalities by Theorem 16. Let $\rho_A$ be the density matrix of quantum shares in a share set $A \subseteq \{1, \ldots, n\}$. Until the end of Sect. 5.4, assume that classical secrets $\mathbf{m}$ are uniformly distributed. From Remark 14 and those exact values of $r_i$ and $t_i$ for $i = 1, \ldots, k$, we see that the Holevo information [32, Section 12.1.1] (quantum counterpart of the mutual information [9]) between $\mathbf{m}$ and $\rho_A$ is

$$I(\mathbf{m}; \rho_A) = \begin{cases} 0 & \text{if } 0 \leq |A| \leq \frac{n}{2}, \\ 2\left(|A| - \frac{n}{2}\right) \log_2 q & \text{if } \frac{n}{2} \leq |A| \leq \frac{n+k}{2}, \\ k \log_2 q & \text{if } \frac{n+k}{2} \leq |A| \leq n. \end{cases}$$

Observe here that one increment of the share size increases the Holevo information by two $\mathbf{F}_q$ symbols. This is in sharp contrast with the classical linear secret sharing [3,7,19,23], because one increment of the share size can increase the mutual information by at most one $\mathbf{F}_q$ symbol in classical linear secret sharing, when each share is one $\mathbf{F}_q$ symbol. Observe also that we have completely determined the classical access structure of this quantum secret sharing scheme.

## 6 Gilbert–Varshamov-type existential condition

In this section, we give a sufficient condition for existence of $C \subset C_{\max} = C_{\max}^{\perp s} \subset C^{\perp s} \subset \mathbf{F}_q^{2n}$, with given parameters.

**Theorem 25** *If positive integers $n$, $k$, $\delta_t$, $\delta_r$ satisfy*

$$\frac{q^{n+k} - q^n}{q^{2n} - 1} \sum_{i=1}^{\delta_r - 1} \binom{n}{i}(q^2 - 1)^i + \frac{q^n - q^{n-k}}{q^{2n} - 1} \sum_{i=1}^{\delta_t - 1} \binom{n}{i}(q^2 - 1)^i < 1, \quad (10)$$

*then there exist $C \subset C_{\max} = C_{\max}^{\perp s} \subset C^{\perp s} \subset \mathbf{F}_q^{2n}$ such that $\dim C = n - k$, $d_s(C^{\perp s}, C_{\max}) \geq \delta_r$ and $d_s(C_{\max}, C) \geq \delta_t$.*

**Proof** The following argument is similar to the proof of Gilbert–Varshamov bound for stabilizer codes [4]. Let $\mathrm{Sp}(q, n)$ be the set of invertible matrices on $\mathbf{F}_q^{2n}$ that does not change the values of the symplectic inner product. Let $A(k)$ be the set of pairs of linear spaces $(V, W)$ such that $\dim V = n - k$ and $V \subset W = W^{\perp s} \subset V^{\perp s} \subset \mathbf{F}_q^{2n}$. For $\mathbf{e} \in \mathbf{F}_q^{2n}$, define $B_V(k, \mathbf{e}) = \{(V, W) \in A(k) : \mathbf{e} \in V^{\perp s} \setminus W\}$ and $B_W(k, \mathbf{e}) = \{(V, W) \in A(k) : \mathbf{e} \in W \setminus V\}$. It is known in mathematics [1, Chapter 7]

- for nonzero $\mathbf{e}_1, \mathbf{e}_2 \in \mathbf{F}_q^{2n}$, there exists $M \in \mathrm{Sp}(q, n)$ such that $M\mathbf{e}_1 = \mathbf{e}_2$, and
- for $(V_1, W_1), (V_2, W_2) \in A(k)$, there exists $M \in \mathrm{Sp}(q, n)$ such that $MV_1 = V_2$ and $MW_1 = W_2$.

For nonzero $\mathbf{e}_1, \mathbf{e}_2 \in \mathbf{F}_q^{2n}$ with $M_1\mathbf{e}_1 = \mathbf{e}_2$ ($M_1 \in \mathrm{Sp}(q, n)$) and some fixed $(V_1, W_1) \in A(k)$, we have

$$\begin{aligned}
&|B_W(k, \mathbf{e}_1)| \\
&= |\{(V, W) \in A(k) : \mathbf{e}_1 \in W \setminus V\}| \\
&= |\{(MV_1, MW_1) : \mathbf{e}_1 \in MW \setminus MV, M \in \mathrm{Sp}(q, n)\}| \\
&= |\{(M_1^{-1}MV_1, M_1^{-1}MW_1) : \mathbf{e}_1 \in M_1^{-1}MW \setminus M_1^{-1}MV, M \in \mathrm{Sp}(q, n)\}| \\
&= |\{(MV_1, MW_1) : M_1\mathbf{e}_1 \in MW \setminus MV, M \in \mathrm{Sp}(q, n)\}| \\
&= |\{(MV_1, MW_1) : \mathbf{e}_2 \in MW \setminus MV, M \in \mathrm{Sp}(q, n)\}| \\
&= |\{(V, W) \in A(k) : \mathbf{e}_2 \in W \setminus V\}| \\
&= |B_W(k, \mathbf{e}_2)|.
\end{aligned}$$

By a similar argument, we also see $|B_V(k, \mathbf{e}_1)| = |B_V(k, \mathbf{e}_2)|$.

For each $(V, W) \in A(k)$, the number of $\mathbf{e}$ such that $\mathbf{e} \in W \setminus V$ is $|W| - |V| = q^n - q^{n-k}$. The number of triples $(\mathbf{e}, V, W)$ such that $\mathbf{0} \neq \mathbf{e} \in W \setminus V$ is

$$\sum_{\mathbf{0} \neq \mathbf{e} \in \mathbf{F}_q^{2n}} |B_W(k, \mathbf{e})| = |A(k)| \times (q^n - q^{n-k}),$$

which implies

$$\frac{|B_W(k, \mathbf{e})|}{|A(k)|} = \frac{q^n - q^{n-k}}{q^{2n} - 1}. \quad (11)$$

Similarly, we have

$$\frac{|B_V(k, \mathbf{e})|}{|A(k)|} = \frac{q^{n+k} - q^n}{q^{2n} - 1}. \quad (12)$$

If there exists $(V, W) \in A(k)$ such that $(V, W) \notin B_V(k, \mathbf{e}_1)$ and $(V, W) \notin B_V(k, \mathbf{e}_2)$ for all $1 \leq \text{swt}(\mathbf{e}_1) \leq \delta_r - 1$ and $1 \leq \text{swt}(\mathbf{e}_2) \leq \delta_t - 1$, then there exists a pair of $(V, W)$ with the desired properties. The number of $\mathbf{e}$ such that $1 \leq \text{swt}(\mathbf{e}) \leq \delta - 1$ is given by

$$\sum_{i=1}^{\delta-1} \binom{n}{i} (q^2 - 1)^i. \tag{13}$$

By combining Eqs. (11), (12) and (13), we see that Eq. (10) is a sufficient condition for ensuring the existence of $(V, W)$ required in Theorem 25. □

We will derive an asymptotic form of Theorem 25.

**Theorem 26** *Let $R$, $\epsilon_t$ and $\epsilon_r$ be nonnegative real numbers $\leq 1$. Define $h_q(x) = -x \log_q x - (1-x) \log_q (1-x)$. For sufficiently large $n$, if*

$$h_q(\epsilon_t) + \epsilon_t \log_q (q^2 - 1) < 1 \text{ and}$$
$$h_q(\epsilon_r) + \epsilon_r \log_q (q^2 - 1) < 1 - R,$$

*then there exist $C \subset C_{\max} \subset C^{\perp_s} \subset \mathbf{F}_q^{2n}$ such that $\dim C = n - \lfloor nR \rfloor$, $d_s(C^{\perp_s}, C_{\max}) \geq \lfloor n\epsilon_r \rfloor$ and $d_s(C_{\max}, C) \geq \lfloor n\epsilon_t \rfloor$.*

***Proof*** Proof can be done by almost the same argument as [30, Section III.C]. □

Theorem 26 has a striking implication that we can construct a secret sharing scheme with roughly 19% of participants being forbidden independently of the size (i.e., $R$ in Theorem 26) of classical secrets for $q = 2$ and large $n$, as $h_2(0.19) + 0.19 \log_2 3 \simeq 1$. Such properties cannot be realized by classical shares.

## 7 Conclusion

In this paper, we considered construction of secret sharing schemes for classical secrets by quantum stabilizer codes and clarified their access structures, that is, qualified and forbidden sets, in terms of underlying quantum stabilizers. We expressed our findings in terms of linear spaces over finite fields associated with the quantum stabilizers and gave sufficient conditions for qualified and forbidden sets in terms of combinatorial parameters of the linear spaces over finite fields. It allowed us to use classical coding theoretic techniques, such as the Gilbert–Varshamov-type argument, and we obtained a sufficient condition for existence of a secret sharing scheme with given parameters. By using that sufficient condition, we demonstrated that there exist infinitely many quantum stabilizers with which associated access structures cannot be realized by any purely classical information processing. We have not thoroughly considered code construction, which is a future research agenda.

**Acknowledgements** The research problem was formulated in a discussion with Diego Ruano during the author's stay at the University of Valladolid, Spain. Without the discussion with him, this paper would not exist. The author also thanks an anonymous reviewer for pointing out a related paper [13].

**Publisher's Note** Springer Nature remains neutral with regard to jurisdictional claims in published maps and institutional affiliations.